# A molecular dynamics test of the Navier-Stokes-Fourier paradigm for compressible gaseous continua

Howard Brenner[1], Nishanth Dongari[2] and Jason M. Reese[2]

[1]Department of Chemical Engineering, Massachusetts Institute of Technology, Cambridge, MA 02139-4307, USA

[2]Department of Mechanical and Aerospace Engineering, University of Strathclyde, Glasgow G1 1XJ, United Kingdom

ABSTRACT

Knudsen's pioneering experimental and theoretical work performed more than a century ago pointed to the fact that the Navier-Stokes-Fourier (NSF) paradigm is inapplicable to compressible gases at Knudsen numbers $(Kn)$ beyond the continuum range, namely to noncontinua. However, in the case of continua, wherein $Kn$ approaches zero asymptotically, it is nevertheless (implicitly) assumed in the literature that the compressible NSF equations remain applicable. Surprisingly, this belief appears never to have been critically tested; rather, most tests of the viability of the NSF equations for continuum flows have, to date, effectively been limited to incompressible fluids, namely liquids. Given that bivelocity hydrodynamic theory has recently raised fundamental questions about the validity of the NSF equations for compressible continuum gas flows, we deemed it worthwhile to test the validity of the NSF paradigm for the case of continua. Although our proposed NSF test does not, itself, depend upon the correctness of the bivelocity model that spawned the test, the latter provided motivation. This Letter furnishes molecular dynamics (MD) simulation evidence showing, contrary to current opinion, that the NSF equations are not, in fact, applicable to compressible gaseous continua, nor, presumably, either to compressible liquids. Importantly, this conclusion regarding NSF's inapplicability to continua, is shown to hold independently of the viability of the no-slip boundary condition applied to fluid continua, thus separating the issue of the correctness of the NSF differential equations from that of the tangential velocity boundary condition to be imposed upon these equations when seeking their solution. Finally, the MD data are shown to be functionally consistent with the bivelocity model that spawned the present study.

**1. Introduction.**
Ever since Knudsen's pioneering publications beginning in 1909 [1,2], fluid mechanicians have known that the compressible Navier-Stokes-Fourier (NSF) equations [3-5] are inaccurate when applied to rarefied gases [6-8]; that is, for *noncontinua*. In those circumstances Boltzmann's gas-kinetic equation is regarded as constituting the fundamental equation governing compressible fluid-mechanical phenomena [9-12]. Despite this knowledge, the NSF equations are currently (albeit implicitly) assumed [3-5] to be valid for *continua*, namely, for vanishingly small

Knudsen numbers $(Kn < 0.001)$[9]. It is this latter belief that our paper aims to critically test using nonequilibrium molecular dynamics (MD) simulations as the probe.

The proposed test, whose precepts are outlined in detail in Ref. [13], involves measuring the radial temperature distribution $T(R)$ in a monatomic gaseous continuum contained within a circular cylinder (radius $= R_o$) possessing a rigid, thermally insulated wall that rotates steadily (angular velocity $= \mathbf{\Omega}$) in an inertial coordinate system. The imposed constraints serve to isolate the gas from its external environment, in the sense of not exchanging any mass, momentum (linear or angular) or energy (internal, potential or kinetic) with its surroundings (outside the rigid wall) during the course of its steady rotation.

According to the solution of the NSF equations for this situation [13] (satisfying a no-slip boundary condition along the cylinder wall), and viewed from the perspective of an observer at rest in this same inertial reference frame, the fluid's mass velocity is readily shown to be one of steady rigid-body rotation, $\mathbf{v}(R) = \mathbf{\Omega} \times \mathbf{R}$ (with $\mathbf{R}$ the position vector relative to a point lying on the axis of rotation) [13]. Concomitantly, the temperature is predicted by the NSF equations to be uniform throughout the rotating gas, independently of radial position $R = |\mathbf{R}|$ (although the pressure and hence density will vary with position as a result of the action of the centrifugal forces). It is this latter belief in the gas's isothermal nature that our simulation aims to test.

Measurements of this temperature distribution via MD simulations were carried out for very small Knudsen numbers, enabling the gas to be regarded as a continuum (namely, $Kn < 0.001\ ca.$), and thereby providing a critical test of the viability of the NSF equations for continua. Specifically, was the temperature found to be nonuniform by an amount exceeding numerical uncertainties in the data, this fact alone would negate current belief in the viability of the NSF equations for compressible gaseous continua. Conversely, was the resulting temperature distribution found to be substantially uniform, this fact would offer credible evidence in support of the applicability of the NSF equations to continua.

As a consequence of the radial symmetries of the temperature, pressure, and density fields under the conditions prevailing above [13], the role of the no-slip boundary condition in the interpretation of the NEMD data proves to be irrelevant. That is, if the fluid slipped at the cylinder wall, the fluid's velocity there would then be $\mathbf{v}'(R_o) = \mathbf{\Omega}' \times \mathbf{R}_o$ rather than $\mathbf{v}(R_o) = \mathbf{\Omega} \times \mathbf{R}_o$ where the difference between the quantity $\mathbf{\Omega}'$ and the actual angular velocity $\mathbf{\Omega}$ of the cylinder's solid wall serves to quantify the slip velocity, $\mathbf{v}_{\text{slip}} = (\mathbf{\Omega} - \mathbf{\Omega}') \times \mathbf{R}_o$. If slip did occur, the velocity at points in the fluid's interior would then be $\mathbf{v}'(R) = \mathbf{\Omega}' \times \mathbf{R}$. The fact that this altered rigid-body fluid velocity differs from that expected in the absence of slip does not, however, impact on the temperature distribution in the case of the NSF equations. Rather, the temperature in the slip case, say $T'(R)$, is predicted on the basis of the solution of the NSF equations for this case, to continue to be uniform throughout the gas (although possibly at a different constant value from that for the no-slip case).

On the other hand, was the temperature found to be nonuniform this would negate the possibility of the NSF equations being correct for continua, irrespective of the extent of the slip occurring at the wall.

## 2. Results of the simulation

Our MD simulations were performed under conditions sufficient to insure that the gas behaved as a true continuum. MD was chosen as the simulation methodology owing to its being deterministic in character, thereby allowing for realistic molecular behavior, i.e., intermolecular attractions, repulsions, movements and scatterings, including realistic interactions with solid surfaces [14].

*Parameters of the simulation:*
We used the open source software program OpenFOAM [15]. It contains a parallelized non-equilibrium MD solver [16-18] that is open source and freely available to download from <www.openfoam.org>. This MD solver was previously rigorously validated for both liquids and gases confined in arbitrary geometries [16-21]. Monatomic Lennard-Jones xenon molecules were simulated [22]. The rigid cylinder wall was taken to be a non-heat-conducting surface, from which gas molecules were reflected at a prescribed velocity corresponding to the wall's angular velocity. The wall was treated as being insulated/frozen, so that no heat transfer took place between gas and its external surroundings through gas-wall interactions; however, the tangential momentum between gas and wall boundary was accommodated, corresponding to the no-slip boundary condition. Owing to the wall's rigidity, no gaseous mass flow occurred through the walls. Collectively, these factors served to isolate the system from its surroundings, namely the region external to the cylinder walls.

Initially, the molecules were spatially distributed throughout the domain of interest with a random Gaussian velocity distribution corresponding to an initially prescribed gas temperature. They were then allowed to relax through collisions until, overall, the system reached a steady state before undertaking measurements. During the relaxation time the system was coupled to a Berendsen thermostat at a temperature of 300 K, which was subsequebtly removed for the latter phase of the simulation. This was implemented to insure that all of our simulations were started with the same prescribed temperature of 300 K. The sampling was performed in the microcanonical ensemble (NVE), consisting of a constant number of atoms, constant volume, and constant energy.

The radius of the cylinder was 7 µm, within which domain 2,000,000 gas molecules were simulated, assuring thereby that the continuum assumption was satisfied. Each case was solved in parallel on 64 cores of the 1100 core HPC facility at the University of Strathclyde. The equations of molecular motion were integrated using a leapfrog scheme [16-18] with a time step of 5 femtoseconds. The actual run time for each case was about 300 hours so as to simulate 10 nanoseconds of problem time after reaching the steady state, i.e., 2 million time samples. The simulation domain was divided into 20 radial bins to measure macroscopic field properties, such as velocity, temperature, and density. Each radial bin has equal volume, so that the widths of each of the bins varied. Hence, in each bin, approximately 200,000 molecules were simulated in order to establish the gas's local macroscopic properties, averaging over 2 million time samples in the steady-state regime so as to minimize numerical errors (e.g., in the case of temperature the maximum error is around ± 0.6K).

To insure that steady-state conditions prevailed during the MD data-recording phase we monitored over time the macroscopic properties of the simulated system, including its overall temperature, local centerline and wall temperatures, kinetic energy, potential energy, total energy, etc. The criterion governing attainment of a steady state was such that the statistical fluctuations of each macroscopic flow property during a specified time interval (around 500,000 time samples in our case) was less than ± 0.002%.

Further details pertaining to the general aspects of our simulation scheme --- in particular to measurements of macroscopic properties in our MD solver --- the reviewer can refer to Refs. [16-21].

*Presentation of data:*

In what follows, we choose to present the data nondimensionally in terms of a 'global' Mach number $Ma = v_o/\bar{c}$, in which $v_o = \Omega R_o$ and $\bar{c} = \sqrt{\gamma \mathcal{R} \bar{T}/M_w}$ is the sonic velocity at the gas's mean temperature, $\bar{T} = (1/2)[T(0) + T(R_o)]$. Here, $\gamma = \hat{c}_p/\hat{c}_v$ is the specific heat ratio (which is 5/3); $\mathcal{R}$ is the universal gas constant, and $M_w$ is the molecular weight. The gas was assumed to obey the ideal-gas law $p = \mathcal{R}\rho T/M_w$, with $\rho$ the density. Thus, for monatomic gases

$$Ma = \sqrt{3(\Omega R_o)^2 M_w/5\mathcal{R}\bar{T}} \ , \tag{1}$$

where, for xenon, $M_w = 83.80$.

With the gas at rest prior to beginning the system's rotation, the pressure and temperature in the cylinder were set at the values $p_0 = 1$ atm and $T_0 = 300°\text{K}$. The Knudsen number for all test cases was less than 0.001 [9]. With $T^* = T(R)/\bar{T}$ the dimensionless temperature, and $R^* = R/R_o$ the dimensionless radial distance from the cylinder axis, Fig. 1 shows the several MD-simulated radial temperature distributions in the cylinder found at each of the six Mach numbers indicated. (The mean temperatures for each case are provided in Table I.)

Table I reports the results corresponding to Fig. 1, expressing the dimensionless temperature difference $\Delta T^* := [T(R_o) - T(0)]/\bar{T}$ between the gas near the cylinder wall and the gas along the cylinder's rotation axis as a function of the Mach number. The uncertainty in the $\Delta T^*$ simulation data was estimated to be $\pm 0.002$ for each Mach number shown.

Table I. Normalized temperature difference in the Xenon gas vs. Mach number

| Mach number | Mean temperature | Normalized temperature difference | Temperature difference |
|---|---|---|---|
| $Ma$ | $\bar{T}$  °K | $\Delta T^* = \dfrac{\Delta T}{\bar{T}}$ | $\Delta T = T(R_o) - T(0)$  °C |
| 0.0999 | 301.21 | 0.00837 | 2.52 |
| 0.1971 | 303.93 | 0.02547 | 7.74 |
| 0.2964 | 307.19 | 0.04681 | 14.38 |
| 0.3881 | 312.57 | 0.07890 | 24.66 |
| 0.4883 | 317.19 | 0.10933 | 34.68 |
| 0.7029 | 341.52 | 0.24311 | 83.03 |

**3. Discussion of the simulation data**

The preceding data show that the temperature is clearly not uniform within the cylinder, increasing from the centerline to the wall. Larger temperature differences are predicted for higher angular velocity values. This temperature difference indicates that the NSF equations do not correctly describe transport phenomena occurring in compressible gaseous continua, a conclusion that holds true independently of the applicability of the no-slip boundary condition at the cylinder walls.

The MD data in Fig. 1 show clearly that that $dT/dR \to 0$ as $R \to 0$, as is obviously required by the assumed radial symmetry of the temperature field. The solution of the NSF equations for the case of an isothermal cylinder surface satisfies this limiting centerline condition *automatically* owing to the trivial fact that $dT/dR = 0$ for *all* $R$, since $T(R) = T(R_o) = \text{const}$ for all $0 \le R \le R_o$. In contrast, the fact that the MD simulation reveals that $dT/dR = 0$ at $R = 0$ constitutes a nontrivial finding.

Though we have used MD to test a situation wherein the wall boundary is treated as insulated, an experiment could equally be designed in which the temperature of the gas at the cylinder surface is simply maintained for all time at some specified value $T(R_o)$, independently of azimuthal position along the cylinder's surface.

**4. Bivelocity theory**

In contrast with the NSF paradigm, the bivelocity paradigm [23] predicts the existence of a *nonuniform* temperature distribution under the circumstances of our simulation [13]. Indeed, it was this knowledge, together with our interest in bivelocity hydrodynamic models [24-28], that provided the impetus for undertaking the simulation.

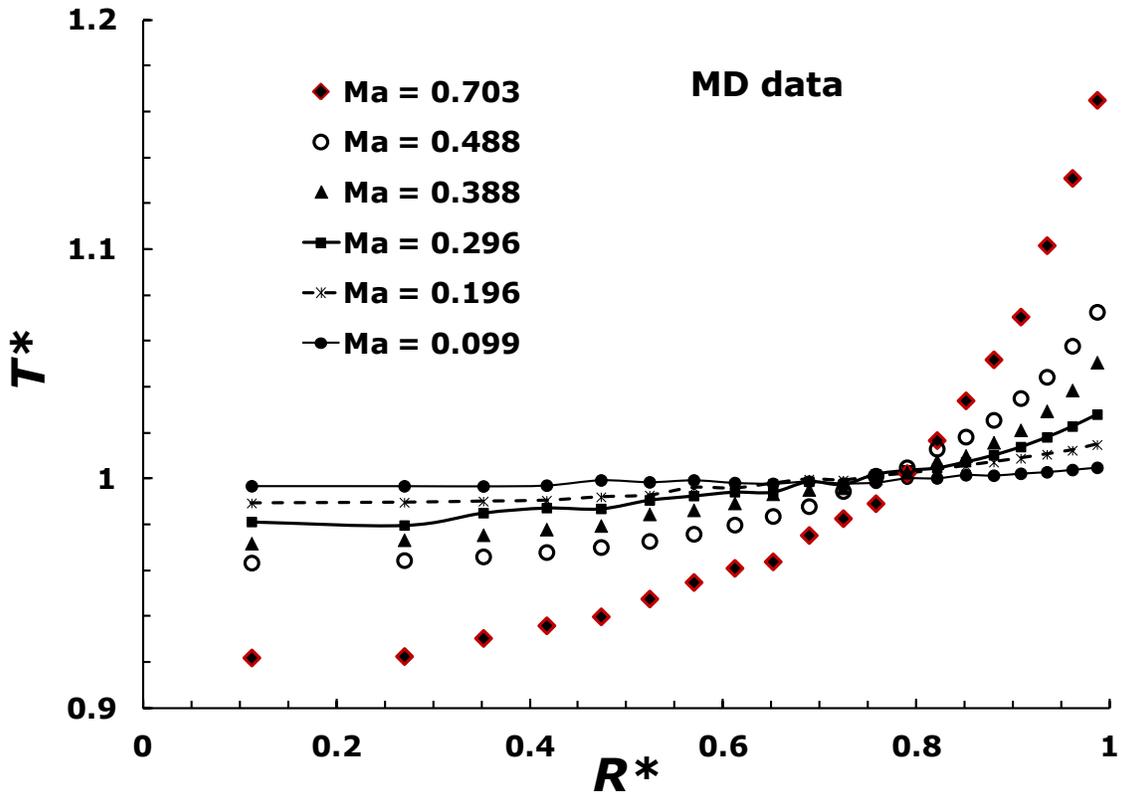

**Fig. 1**. Normalized radial temperature distribution for Xenon gas as a function of Mach number.

For the present isolated rigid-body rotation case, bivelocity theory predicts that [13]

$$\Delta T^* = \frac{1}{3} f(C)(Ma)^2, \qquad (2)$$

in which the parameter $C$ is an $O(1)$ dimensionless phenomenological property of the gas, as defined by its appearance in the constitutive equation

$$\mathbf{j}_v = \frac{C}{Pr} v \nabla \ln \rho , \qquad (3)$$

for the diffuse volume flux [23], wherein $v$ is the kinematic viscosity and $Pr$ the Prandtl number. Values of $C$ close to unity have been cited in the literature (cf. table I of Ref. [28]) based upon both theoretical considerations and experimental rarefied-gas experiments.

Appearing in Eq. (2) is the dimensionless function

$$f(C) = \left[\frac{1}{\gamma}\left(\frac{1}{C}-1\right)+1\right]^{-1}, \tag{4}$$

with $\gamma = 5/3$ in present circumstances. The limiting value of $C = 0$, for which diffusive volume fluxes are absent, makes $f(C) = 0$. From Eq. (3) the value $C = 0$ corresponds to the absence of diffuse volume flux effects, such that $\mathbf{j}_v = \mathbf{0}$, thereby rendering the NSF equations valid [3-5]. This conclusion accords with the fact that on the basis of Eq. (2) $\Delta T^* = 0$ as predicted by the NSF equations [13].

For the purposes of comparing the $\Delta T^*$ simulation values appearing in Table I with the theoretical predictions of bivelocity theory, the constant $C$ may be regarded as a (Mach number-independent) fitting parameter. In this context, the data points given in Table I were plotted against $Ma$ in Fig. 2. The best-fit solid line drawn through the data points corresponds to the empirical equation

$$\Delta T^* = 0.500 Ma^2 \pm 0.002 \tag{5}$$

Based upon use of Eqs. (2) and (4) the preceding expression corresponds to a best-fit bivelocity-theory value of $C = 2.25$ or

$$C = 9/4, \tag{6}$$

as shown by the solid curve in Fig. 2. The functional form of the temperature difference/Mach number correlation predicted by bivelocity theory, namely Eq. (2), is thus seen to provide an excellent fit to the simulation data.

For purposes of comparison with previous experimental/theoretical studies pertaining to the bivelocity model, the dashed curve in Fig. 2 shows the theoretical bivelocity curve for $C = 1$, the latter being the value of the coefficient most frequently cited [28] in connection with *noncontinuum* experiments performed upon compressible gases, as well as with purely theoretical analyses. From Eq. (2) this corresponds to the relation

$$\Delta T^* = 0.333 Ma^2, \tag{7}$$

a result deviating by less than a factor of two from the purely *empirical* simulation-based correlation (5).

## 5. Discussion
*NSF model predictions*:

The results of our simulations served to test the viability of the compressible NSF equations in the very small Knudsen number ($Kn < 0.001$) regime. This critical examination appears never to have been performed rigorously. Owing to the smallness of the Knudsen numbers, the observed nonzero temperature differences $\Delta T \neq 0$, given by the simulation data at the several Mach numbers investigated, cannot be attributed to noncontinuum *Kn* effects. At the gaseous densities encountered in the simulations it would appear appropriate, physically, to hypothesize that xenon obeys the NSF equations for continua. However, our nonisothermal temperature findings set forth in Fig. 1 and Table I serve to show that the compressible NSF equations cannot be regarded as correct for continua.

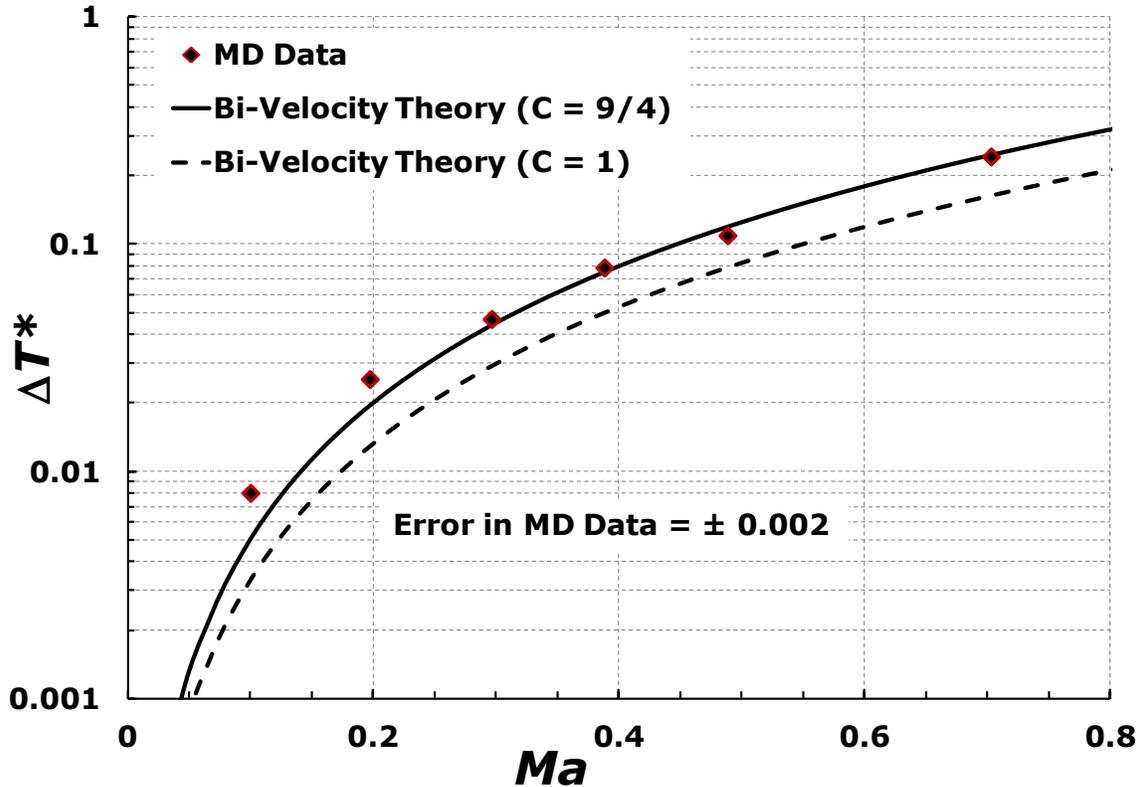

**Fig. 2.** Variation of the normalized temperature difference with Mach number. Using Eqs. (2) and (4), bivelocity theory is compared with the molecular dynamics simulation data for Xenon using a best-fit diffuse volume flux coefficient of $C$ = 9/4. Also shown for comparison is the bivelocity-based curve for the case where $C$ = 1.

Although this conclusion has been demonstrated only for the specific system set forth here, consideration of these findings in conjunction with the known deficiencies of the compressible NSF equations for rarefied gases [1,2,6,8-12] suggesting that problems centered on the realm of validity, if any, of the NSF equations are likely to obtain over the complete range of Knudsen numbers.

In greater detail, our analysis shows that the NSF model for compressible continuum gas flows is accurate, at most, only at small Mach numbers, and then only asymptotically. Hence, the validity of the NSF equations requires that the product of *Kn* and *Ma*, or its equivalent, be small [31]. Moreover, since $Kn = Ma/Re$ (with $Re = R_o v_o / v \equiv \Omega R_o^2 / v$ the Reynolds number), the further requirement that $Kn << 1$, needed for the applicability of the NSF equations to compressible continua, necessitates satisfaction of the strong inequality $Re >> Ma$. As noted from the fluid-mechanical analysis set forth in the original proposal [13], this condition was apparently well satisfied in all of our simulations.

*Bivelocity model predictions*:

The disparity between the bivelocity prediction resulting from using $C=1$ in Eq. (2), and the observed best-fit simulation value of $C=9/4$ does not necessarily exclude the bivelocity model from further consideration as being quantitatively correct. A number of reasons exist for entertaining the possibility that the model may, in fact, eventually prove to be correct despite the disparity between the MD and theoretically-expected *C*-values [28].

In the first place, bivelocity theory is a *linear* theory. For example, it is based, *inter alia*, upon the validity of the linear constitutive equation (3) for the diffuse volume flux. As such, our simulations may have occurred in a parametric range where nonlinear effects were sensible, although likely not dominant. After all, given the intimate connections [23] existing between bivelocity theory's constitutive expressions and those embodied in the Burnett-Boltzmann equations, together with the known presence therein of both nonlinear contributions [7,8] as well as super-Burnett contributions [8,29,30], it may well be that our simulations lay outside the scope of a strictly linear theory.

In addition to the above-cited issues, the bivelocity analysis leading to Eq. (2) [13] invoked a number of hypotheses. One of these assumed that the shear viscosity $\eta$ entering into the calculation of the kinematic viscosity $\upsilon=\eta/\rho$ was constant throughout the analysis, independently of pressure and temperature. The consequences of this assumption entering into the proper *C*-value with which to compare against the simulation-based value, Eq. (6), remains unknown at this time.

These and other possibilities need to be considered and subsequently explored before a final judgment can be rendered as to the viability of the bivelocity model as a possible alternative to the NSF model. To its credit, the bivelocity model is seen to embody [13], at least qualitatively, all of the key features observed during our MD simulation. In any event, the principal findings of this paper lie not with the success or failure of bivelocity hydrodynamics --- for which additional studies are required prior to rendering an authoritative judgment --- but rather with the fact that the simulations demonstrate NSF hydrodynamics to be inadequate in the task of modeling the flow of compressible gaseous continua.

## *6.* Conclusions

Our molecular dynamics simulations have illuminated the limitations of the Navier-Stokes-Fourier equations for compressible gas flows in the *continuum regime*. When added to the fact that NSF equations are also known to be inadequate for noncontinuum compressible gas flows, it follows that the NSF paradigm may not be completely appropriate as the fundamental equations of viscous fluid mechanics, except, possibly, for the case of incompressible fluids. Furthermore, our bivelocity model findings, when considered in conjunction with the success of the bivelocity paradigm in also correlating noncontinuum data (see Table I of Ref. [28]; [23]), suggest the latter as a potential candidate for the fundamental paradigm of viscous fluid mechanics, thus replacing the NSF equations in that role. Further exploration of these themes is required before more definitive conclusions can be drawn.